\documentclass[showpacs,preprintnumbers,amsmath,amssymb,10pt]{revtex4}

\usepackage{graphicx}
\usepackage{dcolumn}
\usepackage{bm}
\usepackage{hyperref}
\usepackage{color}

\begin{document}
\title{Cosmological footprints of loop quantum gravity}

\author{J. Grain}
 \email{grain@apc.univ-paris7.fr}
 \affiliation{%
Laboratoire AstroParticule et Cosmologie\\ 
Universit\'e Paris 7, CNRS, IN2P3\\
10, rue Alice Domon et L\'eonie Duquet, 75205 Paris cedex 13, France\\
{\rm and}\\
Institut d'Astrophysique Spatiale\\ 
Universit\'e Paris-Sud 11, CNRS\\
B\^atiments 120-121, 91405 Orsay cedex, France}%
\author{A. Barrau}%
 \email{aurelien.barrau@cern.ch}
 \affiliation{%
Laboratoire de Physique Subatomique et de Cosmologie\\ 
UJF, INPG, CNRS, IN2P3\\
53, avenue des Martyrs, 38026 Grenoble cedex, France\\
{\rm and}\\
Institut des Hautes Etudes Scientifiques\\
35, route de Chartres, 91440 Bures-sur-Yvettes
}%

\date{\today}

\begin{abstract}

The primordial spectrum of cosmological tensor perturbations is considered as a 
possible probe of quantum gravity effects. Together with string theory, loop quantum 
gravity is one of the most promising frameworks to study quantum effects in the early 
universe. We show that the associated corrections should modify the potential seen by
 gravitational waves during the inflationary amplification. The resulting power 
 spectrum should exhibit a characteristic tilt. 
 This opens a new window for cosmological tests of quantum gravity.

\end{abstract}

\pacs{04.60.Pp, 04.60.Bc, 98.80.Cq, 98.80.Qc}
\keywords{Quantum gravity, quantum cosmology}

\maketitle

\begin{center}
Phys. Rev. Lett. 102, 081301 (2009)
\end{center}

{\it Introduction-} A fully convincing quantum theory of gravity is still missing. Beyond the internal
 theoretical difficulties, experimental probes are extremely difficult to find. The two 
 main paradigms to investigate quantum gravitational effects are unquestionably string 
 theory and Loop Quantum Gravity (LQG). This latter approach is a  promising way to 
 derive a background independent quantum theory of spacetime based only on the 
 experimentally well confirmed principles of general relativity (GR) and quantum mechanics. 
 Introductions can be found in Rovelli's book \cite{rovelli1} and Smolin's review 
 article \cite{smolin1}. Based on the reformulation of GR as a kind of 
 gauge theory obtained by Sen \cite{sen} and Ashtekar \cite{ashtekar1} in the 1980's, 
 LQG is now a language and a dynamical framework which leads to a mathematically 
 coherent description of the physics of quantum spacetime. Important problems are still
 to be solved in LQG, starting with an explicit semi-classical limit recovering
 GR, but the important results obtained both in black hole physics (see, {\it 
 e.g.}, \cite{ashtekar2}) and cosmology (see, {\it e.g.}, \cite{ashtekar3}) urge an 
 experimental test of the theory. There are predictions for observable Planck scale 
 deviations from energy momentum relations \cite{alfaro} that lead to possibly measurable
 effects for and very high energy gamma-ray experiments. Nothing 
 conclusive has however yet emerged from the associated analysis. In this article, we focus on
 a new approach to search for observational probes of LQG \cite{grainmoriond}. It consists in investigating 
 into the details the way the amplification occurs during the inflationary phase when 
 taking into account the modified dispersion relation describing the propagation of 
 gravitons. First, the general framework is described together with the 
 Shr\"odinger-like equation of motion which encodes the LQG corrections in the potential
 term. Then, the primordial tensor power spectrum is 
 explicitly derived and the tilt is given as a function of the LQG Barbero-Immirzi
 parameter for a specific and favoured value of the $n$ parameter. Both de Sitter and
 power-law inflations are considered. Finally, the validity of the overall scheme is
 checked and a few points are made about possible developments.\\

Two basic types of quantum corrections are expected from the Hamiltonian of LQG. These 
corrections arise from inverse powers of the densitized
triad and from the fact that loop quantization is based on holonomies, {\it i.e.}
exponentials of the connection, rather than direct connection components. In the
following article we focus on holonomy corrections as the inverse volume terms were found
to exhibit a fiducial cell dependence and not to be adequate to describe models with a flat
background \cite{ashtekar4}. Unlike studies based on the controversial
superinflation scenario (see, {\it e.g.}, \cite{miel}) we choose in this article to focus on a
conservative approach where the background dynamics is supposed to be described by a standard
de-Sitter or power-law inflation whereas the LQG terms act as corrections on the mode
propagation. This picture is meaningful for at least two reasons. First, at the fundamental
level, because it is highly probable \cite{tsu} (although still questioned \cite{cop}) that 
superinflation cannot solve alone the problems of cosmology and should be considered as an appropriate
way to set the initial conditions for the standard inflation. Our approach holds in this
second phase. Then, at the heuristic level, because in such an intricate
framework, it is very
useful not to mix all the effects. A full LQG treatment of the cosmological evolution is not
yet possible and the aim of this paper is to focus on a well determined effect.
It was shown \cite{bojo1} that the propagation of gravitons in a FLRW 
universe is given, when holonomy corrections (which provide higher order and higher spatial derivative terms) are taken into account, by the 
following equation of motion:
\begin{widetext}
\begin{equation}
	\left[\frac{\partial^2}{\partial\eta^2}+\left(\frac{\sin{(2\gamma\bar\mu\bar{k})}}{\gamma\bar\mu}\right)\frac{\partial}{\partial\eta}-\nabla^2-2\gamma^2{\bar\mu}^2\left(\frac{\bar{p}}{\bar\mu}\frac{\partial\bar\mu}{\partial\bar{p}}\right)\left(\frac{\sin{(\gamma\bar\mu\bar{k})}}{\gamma\bar\mu}\right)^4\right]h^i_a=16\pi{G}S^i_a
	\label{equbojo}
\end{equation}
\end{widetext}
where $\eta$ is the conformal time defined by
$
d\eta=dt/a(t),
$	
$\bar\mu$ is a parameter related to the action of the fundamental Hamiltonian on a lattice state that can be understood as the coordinate size of a loop whose
holonomy is used to quantize the Ashtekar curvature components, $n$ is so that $\bar \mu$ depends on
the triad component through $\bar\mu\sim\bar p ^n$, and $\gamma$ is the
Barbero-Immirzi parameter. The right-hand side of this differential equation 
corresponds to the source term of gravitational radiations. It also receives corrections from holonomies and 
vanishes in the absence of matter. The friction term and the last 
term of the left-hand side are given by the background evolution, as solved in the LQG 
framework. One can then compute the equation of motion for tensor
perturbation modes with $\bar\mu=\left({\bar{p}}/\lambda\right)^{n}$, 
$n\in[-1/2,0]$. The value of $n$ depends on the scheme adopted to quantize 
holonomies. Furthermore \cite{bojo1}, $\bar{p}=a^2(\eta)$ and 
$\lambda$ has to be chosen so that $\bar\mu$ has the dimension of a length.
The exact value of $\lambda$ and its dependence upon $n$ are still under debate. 
For the specific case $n=-1/2$, it was shown \cite{ashtekar2} that 
$\lambda=2\sqrt{3}\pi\gamma\ell^2_{\mathrm{Pl}}$ and it seems quite natural 
to  phenomenologically parametrize $\bar\mu$ by 
$
	\bar\mu\equiv\alpha\ell_{\mathrm{Pl}}\bar{p}^n.
$
Using Eq. (29) to (31) from Bojowald \& Hossain
\cite{bojo1}, the equation of motion (\ref{equbojo}) can be re-written as a function of the
cosmological parameters (the scale factor $a(\eta)$ and the energy density
 of the background $\rho(\eta)$) and of three LQG parameters ($n$, $\alpha$ and
 $\gamma$):
\begin{equation}
\left[\frac{\partial^2}{\partial\eta^2}+\frac{2}{a}\frac{\partial{a}}{\partial\eta}\frac{\partial}{\partial\eta}-\nabla^2-\frac{2n\gamma^2\alpha}{M^2_{\mathrm{Pl}}}\left(\frac{8\pi{G}\rho}{3}\right)^2a^{4+4n}\right]h^i_a=\bar{
S^i_a},
	\label{equ0}
\end{equation}
where we have replace $\ell_{\mathrm{Pl}}$ by $1/M_{\mathrm{Pl}}$ and $\bar{S^i_a}=16\pi GS^i_a$. 
Introducing a new field $\Phi^i_a=a(\eta)h^i_a$, Eq. (\ref{equ0}) now reads:
\begin{equation}
\left[\frac{\partial^2}{\partial\eta^2}-\nabla^2-\frac{\ddot{a}}{a}-\frac{2n\gamma^2\alpha}{M^2_{\mathrm{Pl}}}\left(\frac{8\pi{G}\rho}{3}\right)^2a^{4+4n}\right]\Phi^i_a=a(\eta)\bar{S^i_a},
	\label{equholo}
\end{equation}
where $\ddot{a}$ is the second derivative according to the conformal time $\eta$. Decomposing the 
field into its spatial Fourier modes, 
$
	\Phi(\vec{x})=\displaystyle\int d^3k(2\pi)^{-3}\phi_k\exp{(i\vec{k}\cdot\vec{x})},
$ and using the standard Friedmann equation relating the Hubble constant to the energy 
density of the background, one has to deal with the following 
equation:
\begin{equation}
\left[\frac{\partial^2}{\partial\eta^2}+k^2-\frac{\ddot{a}}{a}-\frac{2n\gamma^2\alpha}{M^2_{\mathrm{Pl}}}\left(\frac{\dot{a}}{a^2}\right)^4a^{4+4n}\right]\phi_k=a(\eta)\bar{S^i_a}.
	\label{equtau}
\end{equation}
It should be underlined that strictly speaking the tree-level Friedman equation
in LQG reads as
\begin{equation}
	\left(\frac{\dot{a}}{a}\right)^2= \frac{8\pi G}{3}\rho\left(1-\frac{\rho}{\rho_{crit}}\right)
	\label{eqfriedlqg}
\end{equation}
where $\rho_{crit}\sim \rho_{Pl}=1/Gl_{Pl}^2$ in \cite{ashtekar4}. Our whole approach
consistently assumes $\rho \ll \rho_{crit}$.

Eq. (\ref{equtau}) is easily interpreted if one remembers the dynamical 
equation for gravitons in a FLRW background without LQG corrections:
\begin{displaymath}
	\left[\frac{\partial^2}{\partial\eta^2}-\nabla^2-\frac{\ddot{a}}{a}\right]\Phi^i_a=16\pi{G}a(\eta)\tilde{S}^i_a,
\end{displaymath}
with $\tilde{S}^i_a$ the source term in GR. Holonomy corrections
appear as a modification of the dispersion relation. This modification is 
encoded in the last term of the left-hand side of Eq. (\ref{equholo}) and depends on the 
dynamics of the universe through the scale factor, on its content through the 
energy density of the background and on the LQG parameters. It can also be
noticed that, in addition to its time dependence through the scale factor, the 
correction term scales, as expected, as
$
	E_{\mathrm{background}}/M_{\mathrm{Pl}}.
$
This equation can be interpreted as a Schr\"odinger equation, $k^2$ being the energy
and $\ddot{a}/a+2n\alpha\gamma^2\left({\dot{a}}/{a^2}\right)^4a^{4+4n}$ being the
potential $V$. In GR, the potential is simply given by $\ddot{a}/a$.\\

{\it Power spectra-} To derive explicitly the power spectrum, we focus on the case $n=-1/2$.
There is a lively debate within the LQG community between two versions 
of the theory and we begin by considering the Ashtekar, Pawlowski and Singh scheme. In this
approach, the choice $n=-1/2$ is necessary (see, {\it e.g.}, \cite{corichi}) to 
ensure that the theory is independent on arbitrary background structures introduced in the 
process of quantization.
As we deal with the amplification of vacuum quantum fluctuations, the source term in Eq. (\ref{equtau}) is set to zero.
For a de Sitter (dS) inflation, the scale factor is given by $a(\eta)=-1/H\eta$ with $\eta<0$. Therefore, the equation of motion for gravitons 
\begin{displaymath}
	\frac{\partial^2\phi_k}{\partial\eta^2}+\left(k^2-\frac{\nu^2-\frac{1}{4}}{\eta^2}\right)\phi_k=0~~;~~\nu^2-\frac{1}{4}=2-\alpha\left(\frac{\gamma{H}}{M_{\mathrm{Pl}}}\right)^2
\label{equhalf}
\end{displaymath}
can be solved by a Linear Combination (LC) of Hankel functions \footnote{We
choose the Hankel functions as the {\it basis} for the solution as they
satisfy the correct Wronskian condition for a standard particle interpretation of 
quantum field theory.}:
\begin{equation}
\phi_k(\eta)=\sqrt{-k\eta}\bigg(A_kH_{\nu}(-k\eta)+B_kH^\dag_{\nu}(-k\eta)\bigg),
\label{eqhankel}
\end{equation}
$A_k$ and $B_k$ being two constants of integration determined
by the initial conditions found by studying the region where the adiabatic vacuum can be defined 
(see, {\it e.g.}, \cite{martin}). In this so-called Bunch-Davies vacuum, the solutions are plane waves
\begin{equation}
	\lim_{k\eta\to-\infty}\phi_k(\eta)=\frac{4\sqrt{\pi}}{M_{\mathrm{Pl}}}\frac{e^{-ik\eta}}{\sqrt{2k}}
	\label{ini-cond}
\end{equation}
whose matching with the previously given general solution allows to determine $A_k$ and $B_k$. 
The choice of the vacuum is clearly a non-trivial question in this framework.
However, to remain consistent with our hypothesis, we assume the de-Sitter background to be a
correct approximation from the beginning of inflation. In this case, the appropriate conditions
will inevitably be met in the remote past , {\it i.e.} when $k\eta\to-\infty$. This will not be
true anymore when considering inverse volume corrections \cite{grain}.
Finally,
the solution is expanded around the high amplification regime, which leads to the following
power spectrum, defined as
$P_{\mathrm{T}}(k)\equiv 2k^3/(\pi^2)\times\left|\phi_{k}(\eta_f)\right|^2/a^2(\eta_f)$
(using the convention of \cite{martin}):
\begin{equation}
	\mathcal{P}_{\mathrm{T}}(k)=A_{\mathrm{T}}k^{3-2\nu}~~;~~A_{\mathrm{T}}=\left(\frac{2^{\nu+3/2}\Gamma(\nu)H}{\pi{M}_{\mathrm{Pl}}}\right)^2.
\end{equation}
It can be seen that LQG corrections do not only modify the normalization of the spectrum but also, and
more importantly, lead to a departure from scale invariance. The tensor index is given by:
\begin{equation}
n_{\mathrm{T}}=3-2\nu\simeq \frac{2\alpha}{3}\left(\frac{\gamma{H}}{M_{\mathrm{Pl}}}\right)^2+\mathcal{O}(H^4/M^4_{\mathrm{Pl}})
\end{equation}
which shows that the spectrum becomes blue for a real-valued Barbero-Immirzi parameter (which is
the more realistic case, as inferred, {\it e.g.}, from the entropy of black holes) and red if
$\gamma$ is imaginary. This can be easily understood at the intuitive level. The amplification
starts at the critical time $\eta_c=-\sqrt{\nu}/k$. If $\gamma\in\mathbb{R}$ this is {\it later}
than in the standard general relativistic case and modes are therefore {\it less} amplified. As the difference
between the value of $\eta_c$ with LQG corrections and without LQG corrections becomes smaller
as $k$ increases, short wavelengths will undergo a smaller suppression. The spectrum will
therefore be blue. 

\begin{figure}[ht]
	\begin{center}
	\includegraphics[scale=0.4]{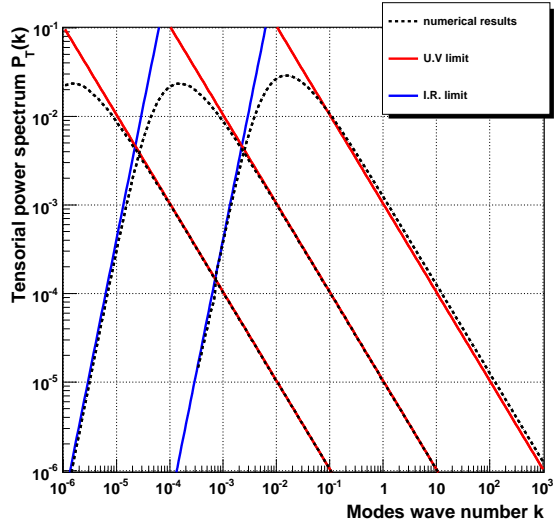}
	\caption{Primordial power spectrum, in Planck units, for power-law inflation with 
	$\beta=-2.5$ and $H_0/M_{\mathrm{Pl}}=10^{-4},~10^{-3}$ and $10^{-2}$ (from the left 
	to the right): dashed lines are for the numerical results, blue lines are for the IR 
	limit and red lines for the UV limit of the power spectrum. 
	The UV approximation coincides with the primordial spectrum obtained without LQG correction.}
	\label{fig-spec}
	\end{center}
\end{figure}

\begin{figure}[ht]
	\begin{center}
	\includegraphics[scale=0.4]{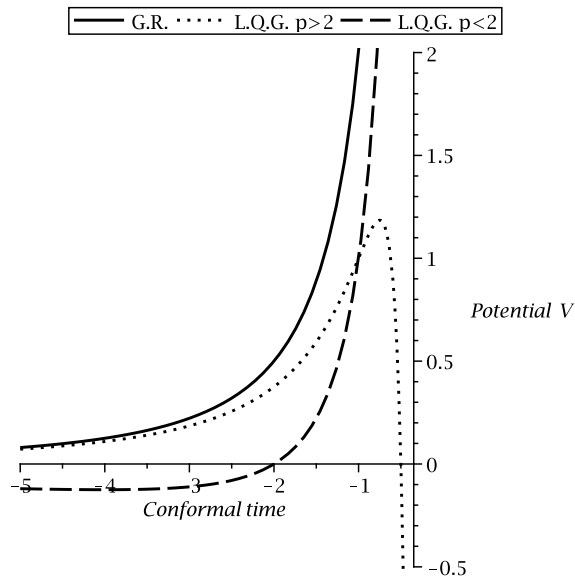}
	\caption{Qualitative shape of the effective potential during inflation
	without LQG corrections (solid line) and with LQG corrections: the dashed
	line is for $p<2$ and the dotted one for $p>2$.}
	\label{pot-fig}
	\end{center} 
\end{figure}

Several cases can be investigated into more details. First, it is worth considering a
power-law inflation. For this inflationary model, the scale factor is given by 
$a(\eta)=\ell_0\left|\eta\right|^{\beta+1}$, with $\beta<-2$. Using
$\delta=\sqrt{\alpha}(\beta+1)^2\gamma H_0/\left|\beta+2\right|M_{\mathrm{Pl}}$, where
 $H_0\equiv\ell^{-1}_0$, the potential term now reads
\begin{equation}
	V(\eta)=\frac{\beta(\beta+1)}{\eta^2}-\frac{\delta^2\left|\beta+2\right|^2}{\left|\eta\right|^{4+2(\beta+1)}}.
\end{equation}	
Because $\beta<-2$, the LQG corrections dominate at the beginning of
inflation and become negligible at the end. Moreover, it can be seen from Fig. \ref{pot-fig}
that adding LQG corrections lowers the effective potential.
As the amplification occurs when $k^2<V(\eta)$, it is obvious from a simple WKB
reasoning that high $k$-modes will not be much affected by LQG corrections whereas low
$k$-modes will be less amplified than in the standard power-law case. LQG
corrections being subdominant at the end of inflation, the mode functions are still
well approximated by a LC of Hankel functions like in Eq. (\ref{eqhankel})
with $\nu=\left|\beta+1/2\right|$. The primordial power spectrum is then given as a function of
$A_k$ and $B_k$ by taking the asymptotic expansion $-k\eta\to0$ of the Hankel functions:
\begin{equation}
\mathcal{P}_{\mathrm{T}}(k)=H^2_0\left(\frac{2^{\nu+\frac{1}{2}}\Gamma(\nu)}{\pi}\right)^2k^{3-2\nu+1}\left|A_k-B_k\right|^2.
\label{spec-pl}
\end{equation}
With the considered potential, the equation of motion for gravitons cannot be solved 
analytically, though the asymptotic limits can be found. Because the LQG term is 
subdominant at the end of inflation, the amplification is effective only in the 
UV regime ({\it i.e.} $k\to\infty$) and the power spectrum should be approximated by the
standard prediction of power-law inflation for high values of the wavenumber. This leads to
\begin{equation}
\left|A_k-B_k\right|^2=\frac{4\pi^2}{kM^2_{\mathrm{Pl}}}.
\end{equation}
Then, it is possible to compute the spectrum in the IR
limit  ({\it i.e.} $k\to0$). The $k-$dependant
solution given in Eq. (\ref{eqhankel}), valid at the end of inflation, is matched to
a solution valid during the entire inflationnary era, but only for $k=0$. By
setting $k=0$ and with the change of variables
$-\eta=\exp{(x/\left|\beta+2\right|)}$ ,
$\phi_0=u(x)\exp(x/2\left|\beta+2\right|)$, the equation of motion takes the
form of a Bessel equation. This allows us to find a solution for the
$(k=0)$-modes function, valid throughout the whole inflationary era, as a LC of Hankel functions.
The
coefficients of this solution are found requiring the Wronskian to be equal to 
$W=-i16\pi/M_{\mathrm{Pl}}$ \footnote{The standard requirement in field theory is that the 
Wronskian should be equal to $-i$. However, $\phi_k$ is already a rescaled quantity and the 
Wronskian condition now reads $W=-i16\pi/M_{\mathrm{Pl}}$. This is in agreement with 
the convention of Martin \& Schwarz and with the normalization of the Bunch-Davies vacuum we used 
to derive the power spectrum for dS inflation.}:
\begin{equation}
\phi_0(\eta)=\frac{i2\pi}{M_{\mathrm{Pl}}}\sqrt{\left|\beta+2\right|}H_\mu\left(\delta\left|\eta\right|^{\left|\beta+2\right|}\right),
\label{kzero}
\end{equation}
with $\mu=\left|\beta+1/2\right|/\left|\beta+2\right|$. Finally, the matching between the two 
solutions is performed at the end of inflation by taking the asymptotic limit of the
solution of 
(\ref{eqhankel}) when $-k\eta\to0$ and the one of solution (\ref{kzero}) when $-\eta\to0$. This leads to:
\begin{equation}
\left|A_k-B_k\right|^2=\left(\frac{2^{\mu-\nu+1}\Gamma(\mu)}{\delta^\mu\Gamma(\nu)}\right)^2\pi^2\left|\beta+2\right|k^{-2\beta-2}.
\end{equation}

This allows us to derive the primordial power spectrum in the IR and UV regimes:
\begin{eqnarray}
\mathcal{P}_{\mathrm{T}}^{(IR)}&=&\left|\beta+2\right|\left(\frac{H_0}{M_{\mathrm{Pl}}}\right)^2\left(\frac{2^{\mu+\frac{5}{2}}\Gamma(\mu)}{\pi\delta^\mu}\right)^2k^3, \\
\mathcal{P}_{\mathrm{T}}^{(UV)}&=&\left(\frac{H_0}{M_{\mathrm{Pl}}}\right)^2\left(2^{\nu+\frac{3}{2}}\Gamma(\nu)\right)^2k^{4+2\beta}.
\end{eqnarray}
In both regimes, the power spectrum is well described by a power-law with a red (blue) spectrum 
in the UV (IR) region. To compute the power spectrum throughout the entire range of 
wavenumbers, the differential equation has been numerically solved with the Bunch-Davies vacuum as initial conditions. The results are 
displayed on Fig. \ref{fig-spec}. This shows the very good agreement between the numerical 
computation and the UV and IR approximations. The value of $\beta$ was
deliberately chosen very small to make the reading easier. As the UV approximation is also the primordial 
spectrum obtained without LQG corrections, it can be seen that adding the LQG correction term 
leads to a suppression of the long wavelength perturbations as compared to the prediction of GR.


The results obtained in the power-law inflation framework can be directly
applied to slow-roll inflation, which
is a realistic scenario in agreement with WMAP data. In particular, the 
asymptotic expansions of Eqs. (15) and (16) are still valid with the appropriate 
replacement of $\beta+1$ by $-1-\epsilon$ \footnote{In the slow-roll approximation, the scale factor is approximated by $a=H^{-1}_0\left|\eta\right|^{-1-\epsilon}$ at the horizon crossing.}. This leads to:

\begin{widetext}
\begin{equation}
\mathcal{P}_{\mathrm{T}}(k)=\left(\frac{4H_c}{\pi}\right)^2\left[1-\epsilon\left(2-\ln{2}+2\ln{\left(-k\eta_c\right)}\right)\right]\Gamma^2\left(\frac{3}{2}+\epsilon\right)k\left|A_k-B_k\right|^2,
\label{spec_sr}
\end{equation}
\end{widetext}
where $H_0$ has been expressed as a function of $\eta_c$ and $H_c$, respectively the conformal 
time and Hubble constant at horizon crossing . A Taylor expansion in $\epsilon$ has 
also been performed.\\

{\it Prospects and validity of the approach-}
In the Ashtekar, Pawlowski and Singh approach considered so far, $n$ must be
equal to -1/2. However, Bojowald has criticized this conclusion (see, {\it e.g.}, \cite{bojo3}
for a very recent discussion) and, for the completeness of our conclusions, it worth 
investigating the case $n>-1/2$ at the qualitative level. For the three inflationary models studied in this article 
(dS, power-law and slow-roll), the potential reads as 
$V(\eta)=\lambda_1/\eta^2-\lambda_2/\left|\eta\right|^p$, where $\lambda_i$ are 
positive constants such that $\lambda_2\ll\lambda_1$ and $p=4-4n(\beta+1)$. Depending 
on the values of $n$ and $\beta$, the exponent of the LQG correction terms $p$ is 
either smaller or greater than 2. (When $n=-1/2$, $p$ is {\it always} smaller than 2).
The shape of the potential is displayed on Fig. \ref{pot-fig} together with the GR 
potential. The case $p<2$ (dashed line) was studied in the previous section, where 
numerical investigations confirmed the na\"\i{v}e WKB-based idea that amplification 
is less effective in the IR regime while roughly the same than in GR in the UV
regime. 
The situation drastically differs when $p>2$ (dotted line) as the LQG term now dominates 
{\it at the end} of inflation (this leads to the drop of the potential when $-\eta\to0$).
With the same WKB reasoning, it is clear that the amplification is now less effective for 
high values of $k$. Moreover, because the potential admits a maximum, the amplification 
totally disappears for $k\to\infty$. Finally, as the potential becomes negative valued, 
the mode functions can start to re-oscillate at the very end of inflation. This 
fundamentally changes the standard picture of the generation of perturbations as the 
temporal coherence of the two-point correlation function is restored {\it before} the 
transition from inflation to the radiation-dominated era. Even with $\beta$
very close to -2, as expected from WMAP data \cite{wmap}, this can happen for $n$ slightly
above -1/2.

It is well known that a complete description of the primordial fluctuations amplified during
inflation requires, in principle, to derive the full quantum statistical properties of the
Bunch-Davies vacuum (the {\it in} vacuum) in terms of quanta defined in the
radiation-dominated era (the {\it out} quanta). In the standard GR case, the quantum
statistical distribution is maximally coherent due to the high enough
amplification of quantum fluctuations ({\it i.e.} when $k/aH\to0$). This high level of
coherence ensures the classical description to be a good approximation at the level of power spectra. In other words tensor fluctuations can be described by a classical, Gaussian distribution with zero mean and a variance proportional to the power spectrum (see \cite{parentani} for a detailed discussion of the link between the apparent classicality and the coherence of the quantum distribution). When LQG corrections are considered, the above statement still holds when $n=-1/2$ simply because LQG corrections are negligible at the end of inflation. When $n>-1/2$, LQG correction are now dominant at the end of inflation and a complete analysis is needed. For super-horizon modes, an asymptotic solution in terms of Bessel functions of the field equation can be found, allowing us to explicitly compute the Bogoliubov coefficients relating the {\it in} vacuum to the {\it out} quanta. In the limit $k/aH\to0$, the two Bogoliubov coefficients are related one to the other by
\begin{equation}
	\beta_k=\alpha^\dag_k\exp{\left(-2ik/aH\right)}.
\end{equation}
Because $k/aH\ll1$, the above-mentioned relation ensures that the distribution is
maximally coherent and the classical description is still a good approximation. In
other words, the classical description valid in a general relativistic framework still
holds when LQG settings are considered {\it if} inflation lasts long enough.\\

{\it Conclusion-} Testing quantum theories of gravity is probably one of the most important challenge of
current fundamental physics. Loop Quantum Gravity has not yet been experimentally probed but it
seems that cosmological observations could allow for a clear signature of LQG effects. This
article opens a new window on quantum gravity by suggesting a way to possibly observe 
holonomy corrections. Although B-mode detection could be
achieved by the Planck satellite and is the main goal of several dedicated experiments for the
forthcoming decade, the signal remains very difficult to extract from the lensing background and
some further refinements are required to quantify the amplitude of the expected LQG effects.
Furthermore, string gaz cosmology also predicts a deviation from scale-invariance
\cite{bran} and the discriminating criteria should be found. Most importantly, the
formalism established in this article should be used for LQG corrections to scalar
perturbations that are just being investigated \cite{bojo2} and could be even more
promising from the observational viewpoint.

\end{document}